\documentclass[manuscript]{aastex}

\slugcomment{Accepted for publication in The Astrophysical Journal}

\shorttitle{Chandra observation of HESS$~$J1640-465}
\shortauthors{Lemiere et al.}

\begin{document}
%
\title{High Resolution X-ray Observations of the Pulsar Wind Nebula\\ 
Associated with the Gamma-ray Source HESS$~$J1640-465}
%
\author{A. Lemiere\altaffilmark{*}, P. Slane}
\affil{Harvard-Smithonian Center For Astrophysics, 60
Garden Street, Cambridge, MA 02138, USA}
\author{B. M. Gaensler}
\affil{Institute of Astronomy, School of Physics, The University of Sydney, NSW 2006,Australia}
\and
\author{S. Murray}
\affil{Harvard-Smithonian Center For Astrophysics, 60 Garden Street, Cambridge, MA 02138, USA}
%
%
\altaffiltext{*}{Now at Institut de Physique Nucleaire, IN2P3/CNRS, 15 rue Georges Clemenceau, 91400 Orsay, France}
%
%
\begin{abstract}
We present a Chandra X-ray observation of the very high energy 
$\gamma$-ray source HESS$~$J1640-465. 
We identify a point source surrounded by a diffuse emission that 
fills the extended object previously detected by XMM Newton 
at the centroid of the HESS source, within the shell of the radio supernova 
remnant (SNR) G338.3-0.0. 
The morphology of the diffuse emission strongly resembles that of 
a pulsar wind nebula (PWN) and extends asymmetrically to the South-West 
of a point-source presented as a potential pulsar. 
The spectrum of the putative pulsar and compact nebula are well-characterized by an 
absorbed power-law model which, for a reasonable $N_{\rm H}$ value 
of $14\times 10^{22} \rm cm^{-2}$, exhibit 
an index of 1.1 and 2.5 respectively,
typical of Vela-like PWNe. 
We demonstrate that, given the H$~$I absorption features observed along the 
line of sight, the SNR and the H$~$II surrounding region are probably connected and lie 
between 8 kpc and 13 kpc. The resulting age of the system is between 
10 and 30 kyr. 
For a 10 kpc distance (also consistent with the X-ray absorption) 
the 2-10 keV X-ray luminosities of the putative pulsar and nebula 
are $L_{\rm PSR} \sim 1.3 \times 10^{33}  d_{10 \rm ~kpc}^{2} \rm erg.s^{-1}$ and 
$L_{\rm PWN} \sim 3.9 \times 10^{33}  d_{10}^{2} \rm erg.s^{-1}$ ($d_{10} = d / 10{\rm~kpc}$). 
Both the flux ratio of $L_{\rm PWN}/L_{\rm PSR} \sim 3.4$  and the total 
luminosity of this system predict a pulsar spin-down power around 
$\dot{E} \sim 4 \times 10^{36} \rm erg~s^{-1}$.  
We finally consider several reasons for the asymmetries observed in the 
PWN morphology and discuss the potential association with the HESS source 
in term of a time-dependent one-zone leptonic model.
\end{abstract}
\keywords{Pulsar: general --- X-rays: individual(\objectname{XMMUJ164045.4-463131, HESS$~$J1640-465}) 
--- ISM: individual (\objectname{G338.3-0.0}) --- supernova remnants}
%
\section{Introduction}
During 2004-2006, H.E.S.S. (High energy stereoscopic system) 
performed a survey of the inner part of the Galaxy where its excellent
capability provided a breakthrough in the field of pulsar wind
nebulae (PWN) study: 
for the first time the morphological structure of many middle-age PWNe 
were resolved in the $\gamma$-ray band. 
A detailed spectral and morphological analysis of the archetypal
example of this population, PSR$~$B1823-13 and its associated TeV nebula 
HESS$~$J1825-137 revealed for the first
time in $\gamma$-rays a steepening of the energy spectrum 
with increasing distance from the pulsar, probably due to the synchrotron 
radiative cooling of the Inverse-Compton (IC) $\gamma$-ray emitting electrons 
during their propagation (Aharonian et al. 2006b). In this scenario, 
the different sizes observed in X-rays and very high energy (VHE) $\gamma$-rays can be 
explained by the different cooling timescales for the radiating 
electron populations (de Jager et al. 2005).  
However, the question of the unusual large size of this PWN candidate 
compared to what the dynamical simulations predict still remains 
(de Jager et al. 2005; de Jager $\&$ Djannati 2008) 
and is difficult to solve. The confusion is even aggravated because of 
the non-detection of the associated supernova shell, 
like it is the case for most of the other middle-aged TeV PWN candidates (Gallant et al. 2007). 

HESS$~$J1640-465 is one of the sources discovered by HESS during
this Galactic survey in 2004-2006 (Aharonian et al. 2006a). 
This source is marginally 
extended with profile close to a Gaussian shape. The spectrum is well fitted by a 
simple power-law with an index of $\sim$ 2.4 and a total integrated flux above 
200 GeV of $2.2 \times 10^{-11} {\rm ~ erg ~ cm}^{-2}{\rm ~ s}^{-1}$. 
As it is shown in Figure 1, the source is spatially coincident with the SNR G338.3-0.0, known 
from radio observations as a $8'$ diameter broken shell with a 
 7 Jy flux at 1GHz  
(Whiteoak $\&$ Green 1996). 
A radio bridge extending to the north of the system is 
visible in the MOST data and coincident 
with the bright H$~$II region G338.4+0.0 (Whiteoak $\&$ Green 1996), 
possibly connected to the shell. 
The fact that the TeV source is center filled and doesn't match 
the shell has favoured the hypothesis that the emission is produced by a PWN so far.        
This field has been successively observed by ASCA (Sugizaki et
al. 2001) and the Swift X-ray telescope (Landi et al. 2006); 
both detected a highly absorbed source inside the shell, 
with compatible positions and non thermal spectra.\\
In 2006, a dedicated XMM-Newton observation of the 
field of view revealed an extended X-ray object 
at the position ($16^{\rm h}40^{\rm m}45^{\rm s}.4$, $-46^{\rm o}31'31''$(J2000))(Funk et al. 2007). 
This extended source (XMMUJ164045.4-463131) exhibits 
a strongly absorbed spectrum with a potential non-thermal spectrum 
(index 
$\sim 1.7$) centered about the VHE source position. 
However, because of the low number of counts of these observations the spectrum was poorly
constrained, and the insufficient angular resolution was not able to resolve 
any compact nebula, nor a pulsar which could confirm 
the hypothesis of a PWN emission. 

We report here on the first Chandra 
observation of this field. We present a detailed data 
analysis and discuss the physical implications of this 
first case of a complete middle-age composite system 
with VHE PWN emission.
\section{OBSERVATIONS AND ANALYSIS}
We report here on X-ray observation of the field around the 
position of HESS$~$J1640-465 taken in May 2007 (Obs ID:7591) using the 
Chandra X-ray Observatory. 
Data were collected with the Advanced CCD Imaging Spectrometer 
(ACIS) in timed exposure, very faint mode, providing a temporal 
resolution of 3.2 s. The target was imaged on I3 chip, with the 
ASCA source position $\sim 2''$ from the optical axis. The other 
ACIS chips activated during this observation were S2, S3, I0, I1, I2.
This imaging system offers an on-axis spatial resolution of 0.8 
arcsec.\\ 
A total of 30 ks of live-time was accumulated. After removing small 
intervals in which data were not recorded, the final exposure 
time for this observation was 26.4 ks. After standard processing 
had been carried out at the Chandra Science Center, we analyzed
the resultant events list using the CIAO(V3.4), FTOOLS(V6.0.4), 
CALDB(v3.3), and XSPEC(V12.2.1) X-ray analysis software packages.
\subsection{Imaging}
Figure 2 shows an exposure-corrected and smoothed image (using a Gaussian 
with $\sigma=2.5''$) in the 2-10 keV energy range, encompassing 
the central part of the field of view of the ACIS-I array. 
To produce this figure, we divided the map of raw counts by the exposure 
map and normalized by the exposure time in order to produce an image 
in units of $photons~cm^{-2}~s^{-1}$. 
The morphology of the X-ray emission imaged here is consistent 
with the one seen by XMM at lower spatial resolution: we confirm 
the presence of a diffuse nebula seen extending $\sim ~ 1'-3'$ around 
the position of XMMUJ164045.4-463131 (Funk et al. 2007).
But at the higher spatial resolution of the Chandra data, some new 
morphological features become apparent, in particular the two compact 
sources S1 and S2 (see Figure 2 and Table 1): S1 is coincident with the brighter 
part of the diffuse emission (North-East) and the other (S2) is placed on the edge, on 
the North-West side of the nebula.\\
Figure 3 shows the radial profile along arcs centered on the 
bright point-source (S1) position along the NE-SW direction.
The surface brightness profile is presented in unit of $count~arcsec^{-2}$. 
All the point sources of the field have been subtracted except S1.
The dashed line represents Chandra's PSF at an energy of 2.5 keV,
shown at the position of S1. It has been obtained by simulating 
a PSF map using CALDBv3.3 and the mkpsf tool and superimposing it on the 
average background level derived from the dashed circular background regions 
shown in Figure 2, then projecting it on the radial profile. 
It is clear from this figure that the extent of the source S1 
is consistent with the point-spread function of the telescope. 
We can see the compact nebula component lying asymmetrically 
from -20' to +70'.  
The inner core of the emission is consistent with a point source surrounded by extended 
structures lying mainly in the South-West direction. The fact that this diffuse 
component steadily fades with increasing distance from S1 supports the
interpretation in which it is the central source of energy of this emission.\\ 
The brightest inner PWN component (called hereafter comp PWN), is $1' \times 0.9'$ in 
area. Figure 2 shows in addition marginal evidence of a tail, 
apparently emanating from S1. It lies within a $5' \times 2'$ area size to the SW of S1, 
with a major axis along the SE-NW line (called Extended PWN
thereafter). We can note that this faint emission has surprisingly a different 
elongation axis from that associated with the inner core of the nebula.\\ 
\begin{table}[!h]
\centering
\begin{tabular}{c c c c c} 
\hline\hline              
Nb & R.A.    & Decl.  & Err R.A & Err Dec \\
   & (J2000) & (J2000)  & (``) & (``)  \\
\hline\hline 
S1  &  16h40m43.52s &  -46d31m35.4s &  0.2 &   0.2  \\
S2  &  16h41m17.61s &  -46d33m26.1s &  0.3 &   0.2   \\ 
\hline\hline 
\end{tabular} 
\label{table:1}
\caption{Corrected coordinates of S1 and S2 point sources.}
\end{table}
S1 and the diffuse emission are not detectable under 2.5 keV, probably due to the 
fact that both spectrum are highly absorbed. Additionaly S1 appears like the hardest source 
in the field and the diffuse emission around it, is smaller and closer at high energy. 
By contrast, S2 is disconnected from the nebula at high energy, it has no detectable 
emission above 5 keV and is relatively bright in the soft band.\\
At this stage, we can compare the X-ray positions of other sources in the field with their 
optical counterparts to improve the nominal Chandra astrometry 
and derive a corrected X-ray position for S1. 
For this we used four X-ray sources in a $0.7'$ radius field of view 
around S1, whose coordinates match with optical objects in the 
USNO-B1.0 (Monet et al.(2003)). 
The corrected X-ray positions for S1 and S2 point sources are given in Table 1.
\subsection{Spectroscopy} 
We have extracted spectra from the data, using the extraction regions 
A,B,C, and the two background regions shown in Figure 2. We defined the pulsar emission as that arising in a 
circle of 1'' in radius centered on the above S1 coordinates (Table
1): we extracted a total of 95 photons.
 For the compact and extended nebula, the extraction regions were point-source 
subtracted. Accounting for 
the local background yields a total of 387 photons 
for the compact nebula and 494 photons for the total extended emission.
The S2 point source with a soft emission was too faint to perform any spectral analysis.\\ 
Spectra from S1 and the nebula were grouped with a 
minimum of 15 counts per spectral channel and fitted using XSPEC 
software in the 2-10 keV energy band. Because of the strong 
absorption, each part of the nebula was fitted together with S1, in order to be 
better constrained: the result of each fit is separated by a double line in Table 2. 
We fitted successively the compact then extended PWN, together with the spectrum of S1, 
using an absorbed power-law (PL) 
model, allowing the common absorbing hydrogen column density to vary.
The absorption being very strong, the fits appeared not very well 
constrained. We restricted the range of $N_{\rm H}$ and tried to 
explore the parameter space.
The confidence contours in the $N_{\rm H}$-$\Gamma$ plane obtained 
from the S1 and CompPWN fits are shown in Figure 5 and 
reveal that S1 is clearly harder than the nebula, even if 
the strong correlation between index and absorption values makes 
difficult to derive definitive numbers 
with these data. 
The spectrum of S1 is shown in Figure 4, it exhibits a hard
spectral index of $\Gamma = 1.1 \pm 0.4$ and 
an unabsorbed flux in the 2-10 keV range of $F_{\rm PSR} = 1.5 \times 10^{-13}{\rm ~ erg ~s}^{-1}{\rm ~ cm}^{-2}$, assuming $N_{\rm H}=14 \times 10^{22} \rm
cm^{-2}$. 
Using the same absorbing hydrogen column density, the compact 
PWN (shown together with S1 spectrum on Figure 4) has a softer spectral index 
of $\Gamma = 2.5 \pm 0.3$ and an unabsorbed flux of $F_{\rm CompPWN}(2-10 keV)= 4.2 \times 10^{-13} 
{\rm ~ erg~s}^{-1}{\rm~cm}^{-2}$, whereas the total extended PWN spectrum is again 
steeper with an index of $\Gamma = 2.7 \pm 0.5$ and has a total unabsorbed 
flux of $F_{\rm ExtPWN} = 4.6 \times 10^{-13} {\rm~erg~s}^{-1}{\rm~cm}^{-2}$. 
 The indices found are marginally in
agreement with the XMM analysis from Funk et al. (2007), but the
absorption appears here much stronger than what was previously derived by
the authors (a factor 1.5-2 lies between the two).\\  
In order to investigate further the possibility of spectral evolution along the 
extended X-ray emission we created also spectra for three regions over
the emission area.  
Again, these regions were fitted simultaneously with absorbed power-laws and with S1: 
four spectra resulted, they are listed in Table 2. 
These results confirm the significant increasing of index value 
of regions away from S1 despite the large uncertainties.  
We found that, for a column density fixed at the value of $14 \times
10^{22} \rm cm^{-2}$, the spectral index varies from  
2.3 $\pm$0.3 to 3.3 $\pm$ 0.5 throughout the PWN. 
Given the areas and positions of the regions, we can conclude that the average spectral index 
increases with distance from S1.
\section{DISCUSSION}
The spectral analysis of the data has shown a good agreement with a power-law shape, 
underlying a possible non-thermal nature of the X-ray emission.
Additionally the radial variation in spectral index that has been observed indicates the presence of a 
pulsar that powers the synchrotron nebula 
and support the picture in which we are observing the continuum hard X-ray emission of 
this object at the position of S1.
Given these elements, we interpret S1 as a putative
pulsar, and the diffuse component surrounding it as a synchrotron
emission from an associated nebula (PWN). 
The ACIS frame time of 3.2 s does not permit a search for the signal of 
a typical rotation-powered pulsar in this observation. Additional 
X-ray and radio obervations will be needed in order to confirm the presence 
of a pulsar on S1 and investigate in detail the spatial PWN component.\\
\begin{table*}
\centering
\begin{tabular}{c c c c c c c}      
\hline\hline              
Region       & Total count   & Model  & $N_{\rm H}$    & $\Gamma$    &
Norm at 1 keV   & $\chi_{\nu}^{2} / \nu$\\
             & [0.5-8]keV &        & (10$^{22} ~ \rm cm^{-2}$)
&  
&     $10^{-4} \rm ph.cm^{-2}.s^{-1}.keV^{-1}$                                  &            \\
\hline\hline
Pulsar  &  95 $\pm$ 9  & PWL    & 19.3 $\pm$ 4.3 & 1.9
$\pm$0.7   &  1.4$\pm$3.6  & 45/56  \\
   (S1)          &               &        & 14  (fixed)      & 1.1
$\pm$0.4   &  0.3$\pm$0.4  & 45/57   \\
CompPWN &  387 $\pm$ 71  & PWL    &.........  & 3.4
$\pm$0.8   & 48$\pm$95 & 45/56   \\
(B-S1-S2)         &               &        & 14  (fixed)      & 2.5
$\pm$0.3   &  10$\pm$7 & 45/57 \\
\hline\hline
Pulsar  &  95 $\pm$ 9  & PWL    & 19.4 $\pm$ 5.4 & 2.0 $\pm$0.8   &  1.6$\pm$0.5  & 216/221  \\
 (S1)            &               &        & 14  (fixed)      & 1.1
$\pm$0.4  &  0.3$\pm$0.4  & 217/222   \\
ExtPWN &  498 $\pm$ 100  & PWL    & ......... & 3.6
$\pm$1.0  & 108$\pm$209 & 216/221   \\
(C-S1-S2)         &               &        & 14  (fixed)      & 2.7
$\pm$0.5  & 21$\pm$23 & 217/222    \\
\hline\hline
Pulsar  &  95 $\pm$ 9  & PWL    & 20.1 $\pm$ 4.2 & 2.1
$\pm$0.7   &  1.8$\pm$4.6  & 198/237  \\
 (S1)            &               &        & 14  (fixed)      & 1.2
$\pm$0.4   &  0.3$\pm$0.4  & 101/238    \\
Reg1  &  181 $\pm$ 17  & PWL    & .....            & 3.2
$\pm$0.8   & 21$\pm$50  & 198/237  \\
 (A - S1)       &               &        & 14  (fixed)      & 2.3
$\pm$0.3   &  3.4$\pm$3.6  & 101/238   \\
Reg2  &  182 $\pm$ 28  & PWL    & .....            & 3.8
$\pm$0.9   & 48$\pm$10  & 198/237 \\
(B-A)         &               &        & 14  (fixed)      & 2.8
$\pm$0.5   &  8.1$\pm$10.1  & 101/238   \\
Reg3  &  178 $\pm$ 61  & PWL    & ....             & 4.4
$\pm$2.4   &  491$\pm$189  & 198/237   \\
(C-B)        &               &        & 14  (fixed)      & 3.2
$\pm$0.5   &  74$\pm$20  & 101/238  \\

\hline
\hline\hline 
\end{tabular}
\caption{Spectral analysis: The background subtracted number of counts and spectra of S1 and different regions of the PWN are summarized in this table.} 
\label{table:4}   
\end{table*}
Figure 6 shows the Chandra image of the nebula together with 
the MOST 843 MHz radio shell of SNR$~$G338.3-0.0. The 
extent of the TeV source HESS$~$J1640-465 is indicated 
by a dashed white circle. This image shows that the putative pulsar 
position is offset from the center of the radio shell and how the 
center of gravity of the X-ray diffuse emission coincides very 
well with the center of the HESS source. A central radio emission 
ten times fainter that the shell lies at the NE side of S1, but it does not 
match S1 or neither the diffuse an asymmetric X-ray nebula extending in the 
SW part of the field.
\subsection{Distance and age of the composite system}
Following the multi-wavelength picture presented in Figure 6, the PWN candidate and
its putative pulsar are probably associated with the broken shell of
SNR G338.3-0.0: this group of objects could then form a composite system.
To complete this picture, we note the presence of the bright
H$~$II regions G338.39+0.16 and G338.45+0.06 extending to the north 
of SNR G338.3-0.0 (Whiteoak $\&$ Green 1996) and also visible in Figure 1. 
(Caswell $\&$ Haynes 1987) have measured the velocities of the 
H$~$II regions using recombination lines near 5GHz with the Parkes 64-m radio telescope, and 
derived velocities at $-29 ~ \rm km.s^{-1}$ and  $-37 ~ \rm km.s^{-1}$. These velocities 
correspond to a near-far kinematic distance solution of 3-4 kpc/12.7-13.7 kpc using the rotation curve 
from (Brand \& Blitz 1993).\\
In the following we present an H$~$I absorption study towards the SNR G338.3-0.0 and the 
closest H$~$II regions 
and use the velocity of the absorbing feature to derive kinematic distances to the 
remnant as well as basic SNR parameters such as radius and age. 
\subsubsection{H$~$I absorption and distance estimation}
Measuring H$~$I absorption features toward radio-bright Galactic supernova 
remnants is a standard technique for establishing kinematic distances to 
these objects. It is basically supported by the fact that cool foreground 
H$~$I absorbs the background continuum emission from an object. So H$~$I data towards 
a continuum source will exhibit absorption features at various velocities 
and will provide bounds on its V$_{\rm LSR}$ (velocity with respect 
to the local standard of rest). The distance can then be calculated from V$_{\rm LSR}$ 
using the Galactic rotation curve.\\
We used here the Southern Galactic Plane Survey (SGPS) of the 21 cm H$~$I spectral 
line emission along the Milky Way 
(McClure-Griffiths et al. 2005).  
All intensities of these data are reported on a main-beam brightness temperature scale, 
T$_{\rm HI}$. Figure 7 (top panel) shows (in full grey) the averaged H$~$I temperature brightness 
profile in the direction of our system. This profile is the average of profiles taken from 
four background regions close to the SNR, which doesn't match any region in which the continuum radio emission is weak,
(each one is integrated on a 0.04 $\times$ 0.04 deg$^2$ area: they are indicated as black squares 
in Figure 8). We see that the matter is mostly at the negative velocities for this 
longitude and the HI brightness temperature is high all over the line of sight.  
Figure 7 (middle and bottom panels) show the subtraction between the HI profile taken from one SNR region 
and HII region (shown as red squares on figure 8) and the averaged background profile discussed before. We see clear absorption features associated with both the SNR shell and the bright overlapping 
H$~$II region detected in almost the whole radial velocity range from +10 km.s$^{-1}$ to -150 km.s$^{-1}$.\\
These features projected in the ($l,b$) plane are strongly correlated with 
the continuum MOST intensity contours all over the line of sight. 
As an example, Figure 8 shows a detailed view of the neutral gas distribution in the local neighborhood 
of the shell, integrated over the velocity range [-25;-35] km.s$^{-1}$, overlapped by green contours 
showing the radio continuum emission shape in this field. 
We see clearly that the H$~$I 
absorption features (the dark features) match the green radio contours, which allowed us to conclude
that they are in physical connection. 
For our special case in which absorption extends to the most extreme negative velocity, 
the object is judged to be located 
beyond the tangent point of the line of sight, around 8 kpc (using the rotation curve from 
Brand \& Blitz (1993)). 
Additionally, an upper limit to the distance can be fixed at 15 kpc by the fact that there 
is no similar feature spatially correlated with the continuum radio emission at positive velocities 
both for the SNR and closest H$~$II regions whereas, for other H$~$II regions like G338.40-0.23, 
absorptions features are observed. 
For the H$~$II regions G338.45+0.06 and G338.39+0.16, these absorption features 
reject the close distance associated with the velocity derived from 
recombination line, and constrain them at a far distance around [12-13.5] kpc 
(given velocity and line-width uncertainties of 
1 $\rm km.s^{-1}$ to ~3 $\rm km.s^{-1}$, and 1 kpc uncertainty due to large 
errors bars in the rotation curve at large distances).
If we consider that the SNR is associated with the H$~$II regions, then the entire 
system would be comprised in this range.  
Additionally, it is interesting to notice that, considering the $N_{\rm HI}$ and H$_2$ 
matter integrated over the total line of sight, it yields a  
total $N_{\rm H}$  
around $15 \times 10^{22} \rm cm^{-2}$, which is 
compatible with the high value derived from the X-ray PWN spectral analysis. 
These results reinforce the physical connection between SNR G338.3-0.0
and the X-ray PWN that we are studying in this paper.\\ 
In the following, we will adopt a 
typical distance of 10 kpc and derive the next relevant quantities with $d_{10} = d/10{\rm~kpc}$. 
\subsubsection{The Host SNR}
We can at this stage, evaluate the physical parameters of the SNR. Using the radio shell 
diameter of $\theta \sim 8'$, the SNR radius is $R_{\rm SNR} \sim 12 d_{10}$ pc. 
No law density cavity has been observed in the HI data around this system. 
It implies that its large size constrains the SNR to a quite advanced step of evolution, so that we can assume 
that the expansion has entered the adiabatic phase (Sedov phase).  
Considering the self-similar
equations (Sedov 1959) with a density between 1 and 10 $\rm cm^{-3}$, we obtain an age between 
$10 ~ d_{10}^{5/2}$ kyr and $30  ~d_{10}^{5/2}$ kyr.  
The SNR shock temperature $T_{\rm S}$ is given by $k T_{\rm S} = 1.8
\times 10^{5} \Big(\frac{R_{\rm SNR_{\rm pc}}}{t_{\rm yrs}} \Big)
^2 $ keV (Sedov 1959), which gives for this case: $k T_{S}$= $0.13 ~d_{10}^{-3}$ keV.    
The corresponding soft thermal emission which, considering the high $N_{\rm H}$ value 
must be highly absorbed, makes it unsurprising that the emission is not detected in these X-ray data.\\
The SNR shell parameters derived below place this system at a middle age,  
which corresponds to the average age of most of the pulsars proposed to be associated 
with HESS PWN candidates (Lemiere et al. 2007). 
\subsection{The PWN and its pulsar}
\subsubsection{Luminosity and Spectrum}
The 2-10 keV X-ray luminosity of S1 that we designated as the 
putative pulsar of the system, is  $L_{\rm X_{\rm PSR}}~\sim~1.3 
\times 10^{33}~d_{10}^{2}{\rm~erg~s}^{-1}$, whereas the total 
X-ray luminosity of the nebula is estimated to be $L_{\rm X_{
\rm ExtPWN}}~\sim~3.9 \times 10^{33}~ d_{10}^{2} {\rm~erg~s}^{-1}$. 
According to Possenti et al. (2002), this suggests a total pulsar spin-down 
power $\dot{E} \sim  4.5 \times 10^{36}$. 
The large nebula-to-pulsar X-ray flux ratio of 
$F_{\rm PWN}/F_{\rm PSR}~=~3.4$ in the 2-10 keV energy band 
gives also an indication that the putative pulsar is energetic 
(Gotthelf 2003; Kargaltsev et al. 2007). 
Given a TeV luminosity of $L_{\rm TeV} = 2.8 \times 10^{35}~d_{10}^{2} 
{\rm~erg~s}^{-1}$, the TeV efficiency is estimated to be 
$L_{\rm TeV} / \dot{E} \approx 10^{-1}$, similar to those of the very 
powerful PWNe around Vela-like pulsars like HESS$~$J1825-137 
(Pavlov et al. 2007).\\
The index of the total PWN is the steepest one of the middle-aged systems studied by 
Chandra so far, whereas the pulsar index is one of the hardest 
(just after J2021+3651 (Hessels et al. 2004). 
Like for many other sources, the power-law seems to become steeper away from the pulsar 
position (Slane et al. 2000; Torii et al. 2000; Warwick et al. 2001; Slane et al. 2004; Li et al. 2005). 
\subsubsection{Size and evolutionary state of the system}
The compact X-ray PWN angular size is approximately $\theta \sim 1'.2$,
thus $R_{\rm CompPWN}= 3.5  d_{10}$ pc. 
Given the TeV source radius of $R_{\rm TeV}= 10 d_{10}$ pc, we obtain 
$R_{\rm Tev} \sim 3 \times  R_{Xray}$.
Similar large factors between X-ray and TeV size values have been observed for 
several middle-age PWN systems like HESS$~$J1825-135 or HESS$~$J1420-607 (Aharonian et al. 2006b,2006c), and can be explained by the different life-time of the 
emitting electrons: the high energy electrons that produce the X-ray emission, 
are burnt-off before reaching the outer regions of the nebula, whereas it is not 
the case for lower energy electrons that produce the TeV emission.\\
Comparing these sizes with the SNR shell  
gives the following ratio: 
$R_{\rm TeV} \sim 75 \% R_{ \rm SNR}$ and $ R_{\rm X} \sim 25 \%  R_{\rm SNR}$. 
The model from Blondin et al. (2001) predicts a typical $R_{\rm PWN}/R_{\rm SNR}$ 
ratio of around 0.25 to 0.3 for such a system.  
The X-ray nebula size appears here in agreement with this model, whereas 
the TeV size (which is closer to the total PWN extension), is larger by a factor of 2. 
A similar result has been previously observed for the prototype middle-aged PWN candidate HESS$~$J1825-137, 
but in this case the SNR shell was unfortunately not visible and its size was only evaluable 
using a Sedov model. 
The observation of a complete composite system such as the one we studied here, gives for the first time 
a firm example of the large percentage of a SNR shell area that can reach a centered filled PWN seen in TeV. 
Future observation of several similar systems with different ages and environments and 
the detection of the associated pulsars will help to disentangle this current mystery. 
 \subsubsection{Morphology}
One of the main characteristics of the system is that the 
putative pulsar position is not centered on the radio shell of the SNR and 
the nebula extension is asymmetric with respect to the putative pulsar position.
Indeed, S1 is shifted to the nebula NE edge, but as 
far as we observe, the extended emission does not have a cometary aspect. 
Additionaly, we see that the nebular symmetry axis is completely misaligned 
with respect to the vector running from the SNR's center to the pulsar position.\\
Assuming a genuine association between the pulsar candidate, 
the PWN and SNR, a possible explanation would be that the pulsar had moved from its
original position because of a high proper motion.  
For a simple geometric assumption based on the offset between the shell center 
and the putative pulsar position S1 ($\delta \theta \sim 2'$), assuming that the 
center corresponds to the SN site, we obtain a projected space velocity of 
$V_{\rm PSR} = 500 ~d_{10}^{-1.5}~\rm km.s^{-1}$ in the plane of the sky, 
which is fast but not unreasonable for a pulsar.
But for such pulsar space velocity, nebulae develop bow shocks as a result of ram 
pressure confinement through the surrounding medium (van der Swaluw et al. 2001; Gaensler et al.
2003).
In our case, the diffuse emission is misaligned with the potential direction of motion and forms 
a very broad region behind the pulsar with significant uncollimated 
diffuse X-ray emission. Additionally,  
the observed TeV emission is centered on the X-ray nebula and has a symmetrical 
Gaussian shape. 
It seems that the pulsar can not reasonably come from the actual SNR's geometric center.\\ 
Another possibility is that the explosion site did not coincide with 
SNR's geometric center. The blast center and geometric 
center of an SNR can be different in the presence of density gradients or 
space velocity of the progenitor star (Gvaramadze 2002). 
 Additionaly, the fact that the system is in an evolved state, in which the reverse 
shock from the surrounding SNR has propagated back inward and probably collided with 
the PWN (Reynolds $\&$ Chevalier 1984), can add more complexity to the system and 
shift the nebula from the pulsar position in an other direction. 
Since the $^{12}\rm CO$ data available from the Survey of Dame et al. (1986) 
are not enough spatially resolved to allow any final conclusion, a dedicated CO 
observation of the field will be needed as a next step of this work. 
We can already conclude that, given the complex kinematics and distribution 
of the molecular and neutral gas in the vicinity of the SNR, it appears unreasonable to ascribe 
the blast center to the geometric center. The dominant cause of this situation 
probably results from multiple interactions with the local ISM. 
\subsection{Connection with HESS$~$J1640-465}
In this section we test the possible association between the TeV and X-ray emission 
whithin a very simple PWN model. In this preliminary study we check that we can reproduce the 
data within a simple picture, using quantities derived in this paper as pulsar age 
and spin-down power, together with reasonable input parameters. 
We obtain an estimation of the magnetic field and minimum energy 
needed to reproduce the spectral data.\\
The common picture describes pulsars as dissipating their rotational energy 
during their life-time by powering a relativistic magnetized wind of particles. 
This wind decelerates until it reaches the termination shock. Beyond this shock, 
particles flow until they radiate all their energy by emitting photons 
over the multi-wavelength broadband spectrum to form what we call the 
pulsar wind nebula (PWN).  
We consider a simple model of electrons powered by the pulsar and accelerated
at the shock radius, then subsequently injected in the nebula. 
The evolution of the pulsar spin-down power with time is described by:
\begin{equation}
\label{equa::L_pulsar}
\dot{E}(t)=\frac{\dot{E_o}}{(1 + \frac{t}{\tau_o})^{\frac{b+1}{b-1}}} \approx \Bigg\{
                                        \begin{array}{ll}
                                           \dot{E_o} & \qquad (t<\tau_o)\\
                                           \dot{E_o}(t/\tau_o)^{-\frac{b+1}{b-1}} & \qquad (t>\tau_o)\\
                                          \end{array} 
\end{equation}
with the pulsar characteristic slowing-down time $\tau_o$ fixed at a typical value of
500 yrs, and a braking index of $b$=3 implied by the dipole rotator model. 
The injection spectrum is defined by a simple power-law:
\begin{equation}
\frac{dN}{dE}= A E^{-s}
\end{equation}
with the index $s=2.4$ (typical of Fermi acceleration) and a normalization $A$ determined by the upper
and lower electron energy limit and the condition that the
integrated spectrum energy does not exceed $1/ (\sigma +1) ~ \dot{E}$,
where $\sigma$ is the assumed magnetization parameter of the wind defined by 
$\sigma =   \dot{ E}_{\rm B} /  \dot{ E}_{\rm e}$ (Kennel $\&$
Coroniti 1984). 
Here $\dot{E}_{\rm e}$ and $\dot{E}_{\rm B}$ are respectively the
spin-down power split between particles 
and field ($\dot{ E} =  \dot{ E}_{\rm e}~ + ~ \dot{ E}_{\rm B}$). 
The lower limit energy of the electron spectrum is one of the parameters of the fit,
whereas the upper limit energy is set
by the condition that the particle gyroradii do not exceed the
termination shock radius (de Jager $\&$ Harding 1992):
\begin{equation}
\label{Cond_Emax}
 \rho_{\rm L}=\frac{ E }{e  B(r_{\rm s})} << r_{\rm s}
\end{equation}
Where $ r_{\rm s}$ is the shock radius, $\rho_{\rm L}$ the gyroradius, and  
$B(r_{\rm s})$ the magnetic field at the shock radius.  
The magnetic field at the shock radius is expressed by the 
equilibrium of wind and magnetic pressure at the termination shock, as written by  
Kennel $\&$ Coroniti (1984). 
The maximum energy of the injection spectrum at the shock radius at each time $t$ is finally written as:
\begin{equation}
\label{Emax_final}
 E_{\rm max}(t) = \Big(\frac{\sigma}{1 + \sigma}\Big)^{\frac{1}{2}} e \Big(\frac{ \dot{ E}(t)}{c}\Big)^{\frac{1}{2}}
\end{equation}
Radiative losses are dominated by synchrotron losses: 
$-\frac{dE}{dt} \propto B(t)^{2}.E^{2}$.  
We neglect the spatial evolution of the magnetic field in the nebula and 
consider its evolution with time. We assume that $B$ globally decreases
with time because of the adiabatic losses due to the nebula expansion 
and we use a simple power-law shape with 
a characteristic timescale equal to the pulsar slowing-down time ($\tau_{\rm o}$):
\begin{equation}
\label{B}
B(t) =  \frac{B_{\rm o}}{1+(\frac{t}{\tau_{\rm o}})^{\alpha}}
\end{equation}
where $B_{\rm o}$ is the magnetic field strength at $t=0$ and $\alpha$ a parameter of the model. 
Non-thermal emission from radio to X-ray comes from
synchrotron emission and $\gamma$-ray are produced by Inverse Compton (IC) 
upscattering of cosmic microwave background photons (CMB) by the high
energy electrons. We have computed the IC contribution taking into
account the 
contribution of star light and IR dust: the density values used 
(STAR LIGHT= 0.9 $\rm eV.cm^{-3}$ and DUST IR = 0.3 $\rm eV.cm^{-3}$) 
come from the GALPROP code ($R_o \sim 5 \rm kpc$; Moskalenko et
al. 2006). 
\subsubsection{Result and Discussion}
The SED presented in Figure 9 shows X-ray and VHE gamma-ray spectra reproduced 
by this simple one-zone model using the set of parameters 
summarized in Table 3. A parameter space exploration done by hand shows that 
given the imput of the model, magnetic field and minimum energy are rather strongly 
constained ($\pm 0.5 \mu$G for B, and $\pm$ few GeV for $E_{\rm min}$). 
We can see on the figure that the synchrotron cooling break has already passed the 
X-ray and TeV bands and lies around  $E_b^{\rm synch} \sim 10$ eV for the
synchrotron emission (far from the Chandra energy range) and around
$E_b^{\rm IC} \sim 10^{11}$ eV in the IC bump (under the HESS energy 
threshold). 
\begin{table}[htb]
\centering
\label{table:5}
\begin{tabular}{c  c}
\hline\hline
Parameter & value \\      
\hline\hline
T (age)        &    15 kyr\\
$\dot{E}$     &    $4 \times 10^{36} {\rm~erg~s}^{-1}$\\
$\sigma$      &     $10^{-3}$   \\
\hline
$E_{\rm min}$  &     50 GeV\\
B(T)          &      6$\mu$G       \\
$\alpha$      &      0.45\\
\hline\hline    
\end{tabular}
\caption{Set of parameters used in Figure 9: the three first parameters (pulsar age, pulsar spin-down power and sigma) are fixed in the model, the others (minimum energy and magnetic field) have been fitted to reproduce the data.}
\end{table}
\\
The high spin-down power 
used to derive the upper energy of the injected electron
spectrum is needed to reproduce the flux and index 
in the X-ray band. Using a smaller $\dot{E}$ would decrease the
electron upper limit energy which would have the effect of a prematured cut-off 
in the synchrotron spectrum. The minimum energy value is
strongly constrained by the flux level given the age and power evolution of
the pulsar. Its fit value looks reasonable given the known properties
of the pulsar wind, and it constrains the radio emission
to a low flux which is in agreement with the non-detection of any
radio nebula in the field. 
The magnetic field found is typical of such middle-age Vela like objects, 
the decreasing index corresponds to a field which varies from an initial
value of 35 $\mu$G to an actual value of 6 $\mu$G (15 kyr later).  
The shape of magnetic field evolution can affect the spectrum index, 
but the low degree of details in VHE
resolution of the source (because of HESS's poor psf), make such a precise shape
analysis non pertinent in this paper. 
Finally, the VHE flux can be reached using the CMBR alone, but in order to reproduce the 
spectral index, we need the addition of the dust and visible target
components.\\ 
Given the reasonable values of parameters needed to reproduce the 
Chandra and HESS spectra and their relatively good compatibility
with the known characteristics of the middle-aged Vela-like pulsars and PWNe, 
the interpretation of the X-ray and VHE gamma-ray emission as originating 
from a pulsar-driven nebula appears secure. 
\section{Conclusion}
This high-resolution Chandra ACIS observation of HESS$~$J1640-165 inside the shell G338.3-0.0, 
has allowed us to spatially resolve a PWN and its putative pulsar, and perform a spectral analysis. 
We found that the spectra are both highly absorbed and non-thermal, and  
the indices and luminosities resemble those of typical Vela-like pulsar systems.  
Using H$~$I absorption features, we were able to constrain the
distance of the shell between 8.5 and 13 kpc, this distance range being compatible 
with the large $N_{\rm H}$ value derived
from X-ray spectral analysis, we concluded that it is likely
associated with the shell G338.3-0.0. 
We finally discussed a scenario in which the $\gamma$-rays originate
from the IC scattering and X-ray from the synchrotron emission of the 
electron populations in the nebula, and 
show that the VHE source HESS$~$J1640-165 is likely a PWN,
probably associated with a Vela-like pulsar of $\dot{E} =  4 
\times 10^{36} \rm erg.s^{-1}$. 
If this interpretation is true, this object is a nice example of a middle-aged composite system where 
the total PWN extension (traced by the TeV emission) fills almost 75 percent 
of the associated SNR shell area, which is quite surprising given that dynamical simulations 
currently predict a 25 percent.
Future search for similar cases will be useful in order to confirm or not the generality of 
this phenomenon and study its causes in detail. 
In this context, the detection of the associated pulsars will be very helpful. 
A next step of this study could be the high resolution radio imaging of the source 
and the search for pulsed emission originating from this putative pulsar position.
\acknowledgments 
The authors would like to acknowledge the support of their host institutions.
B.M.G acknowledges the support of the Australian Research Council through a Federation Fellowship 
(grant FF0561298). POS acknowledges partial support from NASA contract NAS8-03060.

\begin{figure}[htb]
\begin{center}
\includegraphics[width=0.85\textwidth]{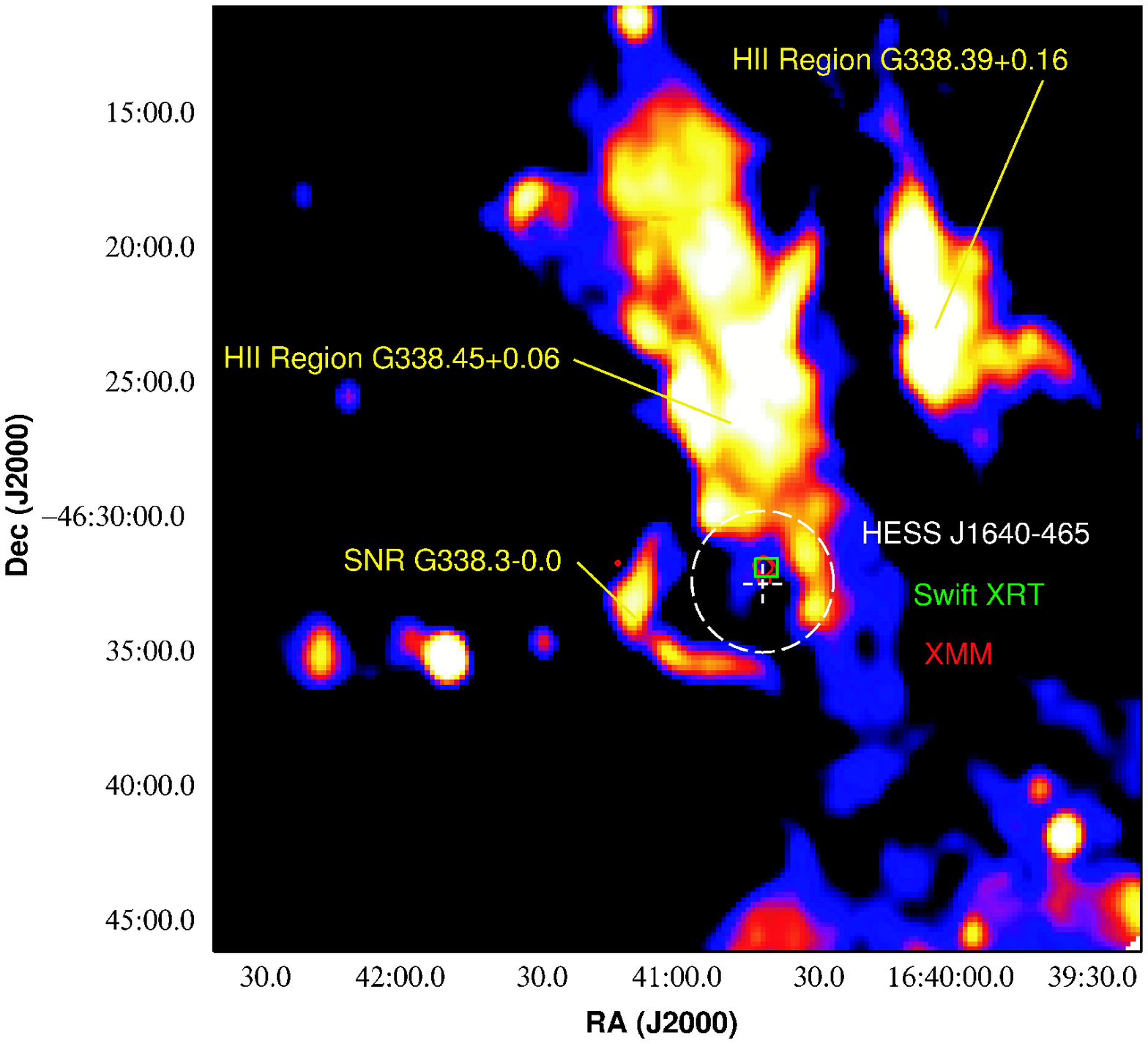}
\caption{\textit{Radio view (MOST 843 MHz) of the field of
G338.3-0.0 in color scale. Swift-XRT point source position is indicated as a green box 
(Landi et al. 2006), and the white plus symbol and circle indicate the position and extension of the HESS source HESS J1640-146. 
The XMM-Newton source (XMMUJ164045.4-463131) extension is also indicated: the 
80\% and 60\% contours of MOS1 and MOS2 count maps are shown in red.}}
\end{center}
\end{figure}

\begin{figure}
\begin{center}
\includegraphics[width=0.80\textwidth]{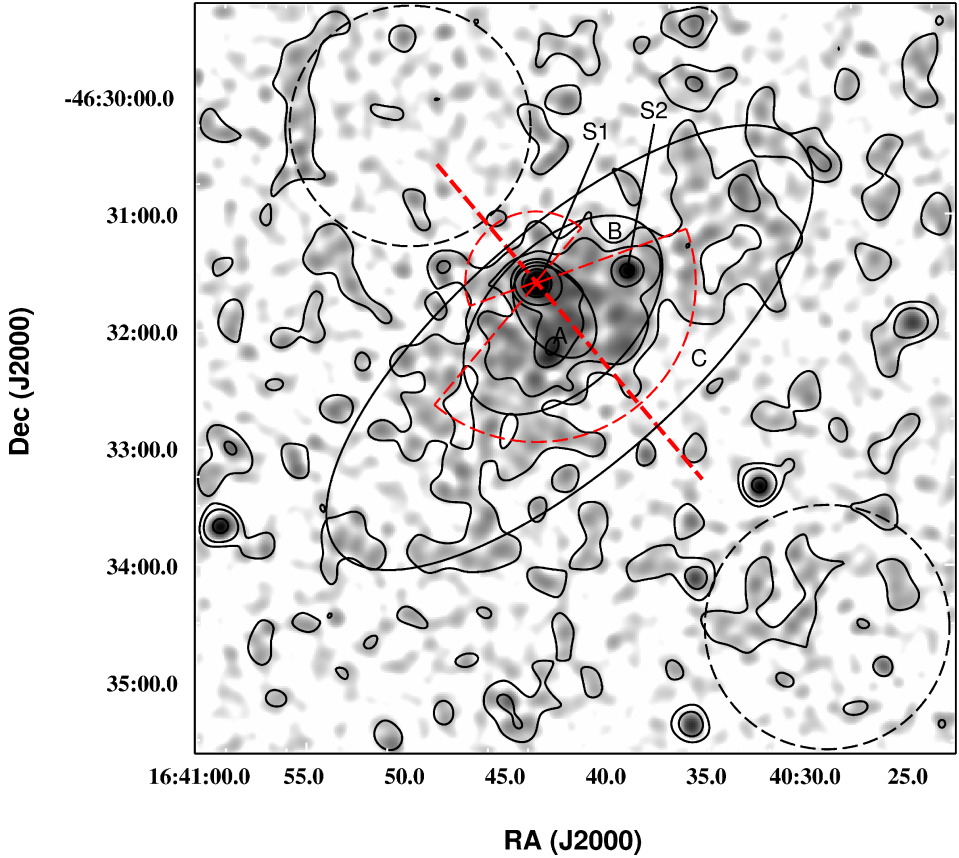}
\caption{\textit{Chandra [2-10 keV] broad-band X-ray image of the field surrounding 
XMMUJ164045.4-463131 from the ACIS-I chips. The image was
exposure-corrected and smoothed with a Gaussian width $\sigma$=2''.5 to highlight 
the extended diffuse emission. Black contours are plotted at levels of 12$\%$, 20$\%$, 68$\%$, 90$\%$, and 
95$\%$ of the peak intensity. The location of the two point-sources 
S1 and S2 are indicated (see Table 1). The three solid black ellipses 
represent the regions A, B and C which encompass the different 
components of the nebula and will be used as 
extraction regions to measure the PWN spectra. The red geometric features show the 
axis and regions from which the radial brightness profile (Figure 3) has been 
computed. Background regions used to derive radial profile and spectra are indicated as dashed black circles.}}
\end{center}
\end{figure}

\begin{figure}
\begin{center}
\includegraphics[width=0.75\textwidth]{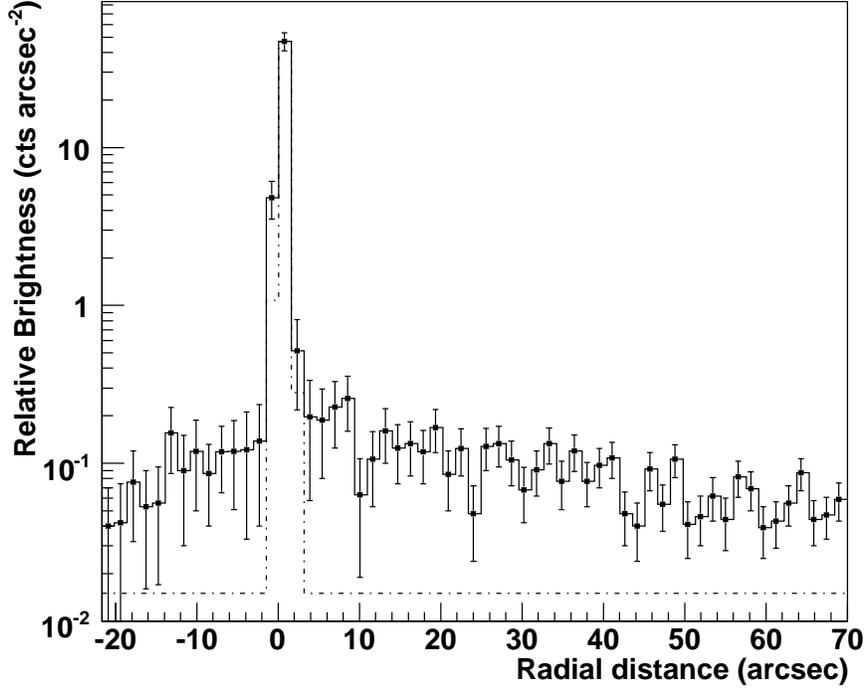}
\caption{\textit{Chandra radial surface brightness profile as a function of increasing distance
from S1 to the SW (positive) and to the NE (negative) (we divided the regions shown in Figure 1 in 65 arcs 
(1''.55 width each)). The profile is shown between $-20$ 
to $+70$ arcsec, and gives the detail of nebula's inner core brightness 
evolution.  We show the  PSF of the telescope determined at 2.5 keV 
(dashed line), normalized to the peak of emission and scaled vertically 
to match the data. S2 has been previously extracted from the data.}}
\end{center}
\end{figure}

\begin{figure}[!h]
\begin{center}
\includegraphics[scale=0.60, angle=-90]{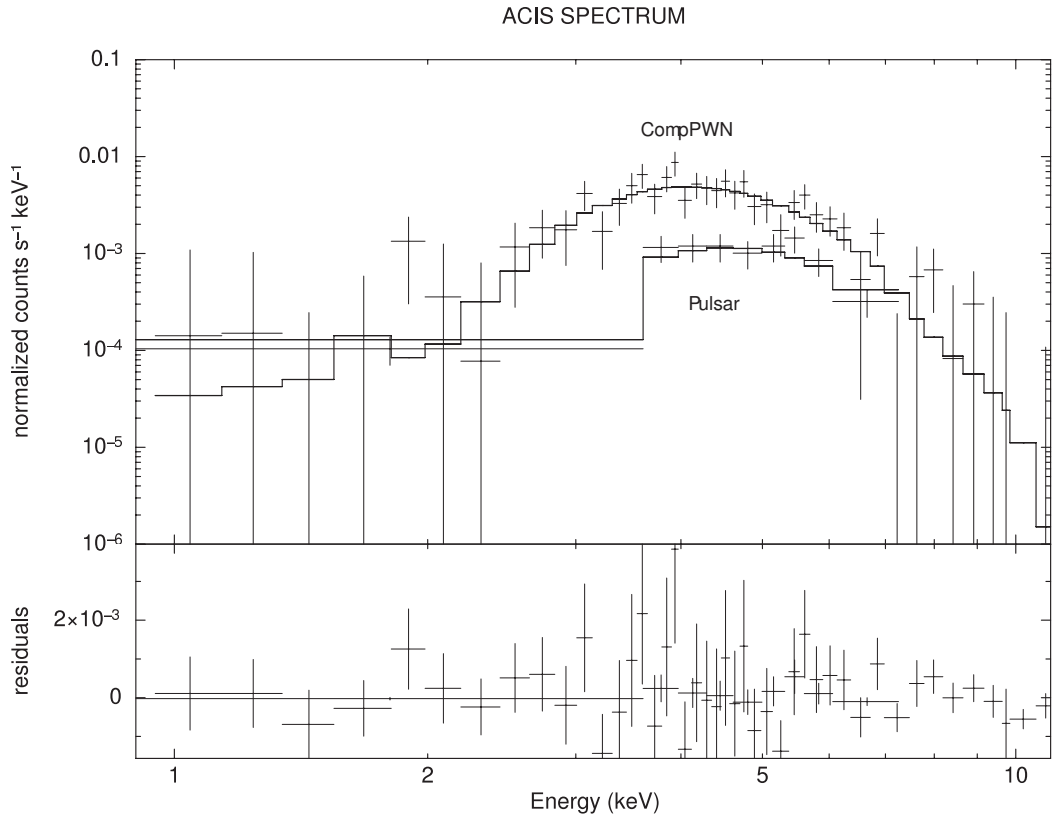}
\caption{\textit{Chandra ACIS spectrum of the putative pulsar (S1) and its compact nebula 
(CompPWN), both fitted with an absorbed power-law model with a linked $N_{\rm H}$. 
Residuals from the best-fit model (summarized in Table 2) are shown in the bottom panel.}}
\end{center}
\end{figure}

\begin{figure}[!h]
\begin{center}
\includegraphics[scale=0.80,angle=-0]{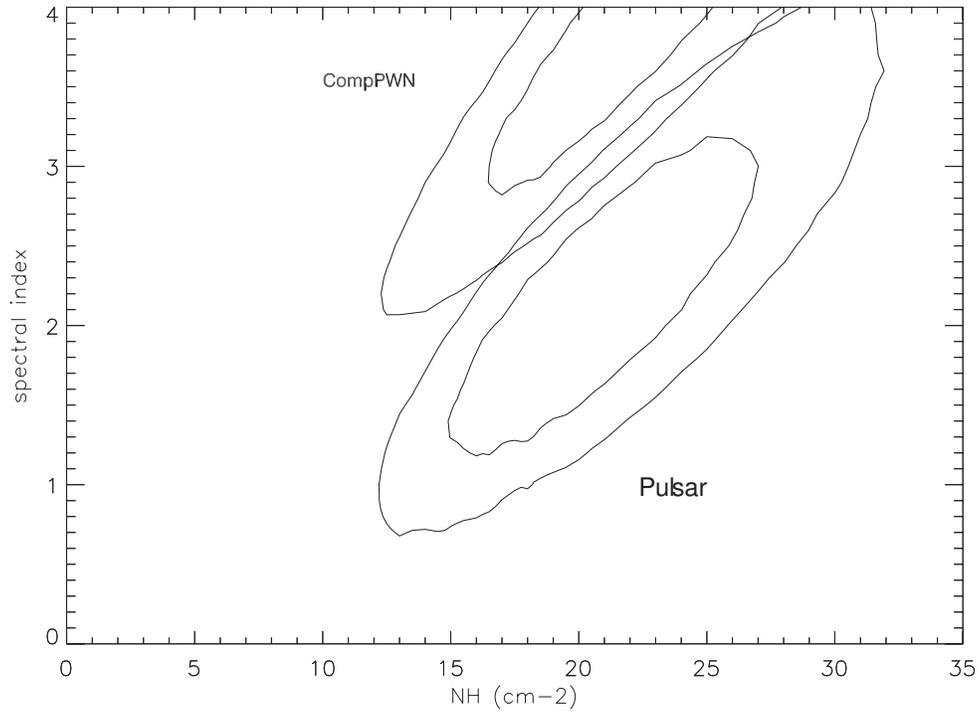}
\caption{\textit{Confidence contours (68$\%$ and 90$\%$) in the $N_{\rm H}$-$\Gamma$ plan 
for the PL fit to the pulsar candidate (S1) and the compact nebula
component (Comp PWN) spectra with a common $N_{\rm H}$ parameter. The contours are obtained with the PL normalization fitted 
at each point of the grid.}}
\end{center}
\end{figure}

\begin{figure}[htb]
\begin{center} 
\includegraphics[scale=0.75, angle=0]{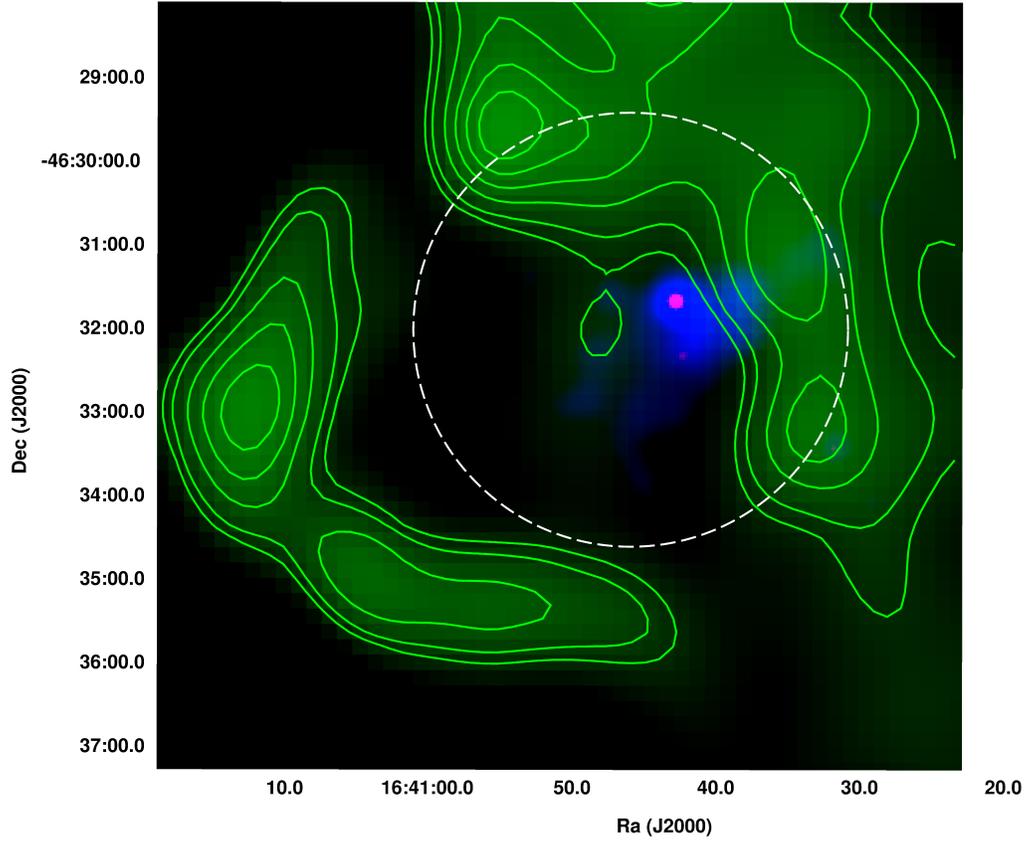}
\caption{\textit{Multi-wavelength view of the field of G338.3-0.0: Chandra emission 
is shown in red ([4.5-8] keV band, smoothing angle of 1''.5) and blue ([2.5-4.5] keV band, smoothing angle of 3.5'') with a 
log intensity scale, and the radio emission from 843 MHz MOST is shown in green (we supperimposed radio data contours on the shell for a better visibility). 
The intrinsic extent of HESS$~$J1640-465 is shown as a white dashed circle.}}
\end{center}
\end{figure}

\begin{figure}
\begin{center}
\includegraphics[width=0.65\textwidth]{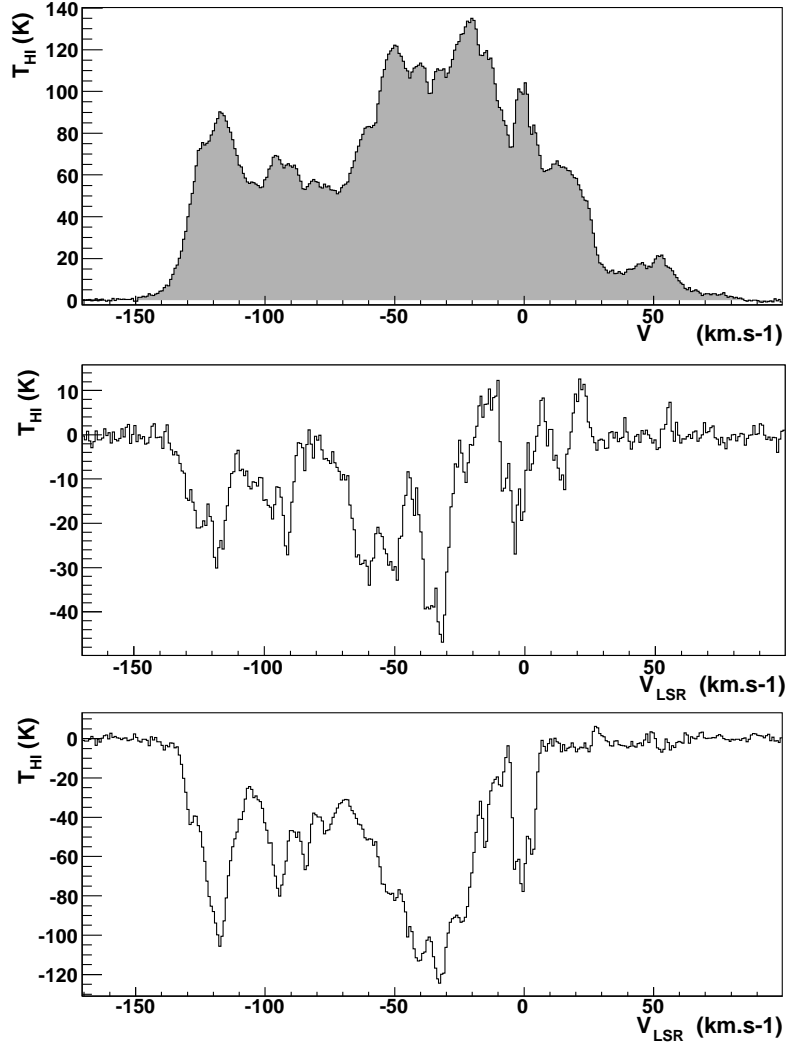}
\caption{\textit{Top panel:} H$~$I 21 cm line temperature brightness intensity profile (K) as a
function of radial velocity ($km.s^{-1}$). This profile shows the averaged background H$~$I 
emission spectrum near the SNR G338.3-0.0. Four individual spectra have been used to 
create this averaged spectrum: each of them has been integrated in the (l,b) plane over a 
0.04 $\times$ 0.04 deg$^2$ area (shown in Figure 8). \textit{Bottom panels:} Absorption 
spectra in the line of sight of the SNR shell, and H$~$II region G338.40-0.23. 
These spectra have been computed by subtracting the averaged background spectum, 
from the spectrum taken from the SNR shell and HII region (Figure 8). We can see clear 
absorption features all over the line of sight for both objects.}
\end{center}
\end{figure}
\begin{figure}
\begin{center}
\includegraphics[scale=0.60, angle=0]{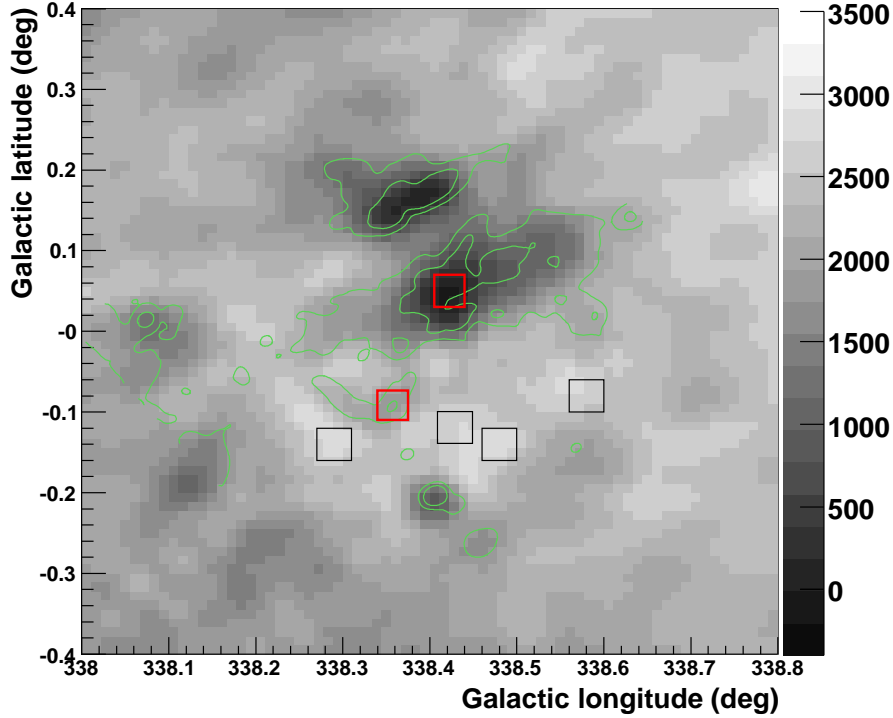}
\caption{ 
\textit{This Figure shows the H$~$I temperature brightness ($\rm \int T_{\rm HI}$ (K)) integrated 
over the velocity range [-25;-35] km.s$^{-1}$, in the field of SNR G338.3-0.0.  
The map is superimposed with the MOST radio intensity contours at 843 MHZ in green. 
We see a clear correlation between the H$~$I absorption features (the darkest regions of the map) 
and the contours. Additionally, four black squares show the background regions used to compute the Figure 7 
top panel averaged profile, and two red squares show the SNR and H$~$II regions used to compute 
the absorbed spectra shown in Figure 7.}}
\end{center}
\end{figure}

\begin{figure}
\begin{center}
\includegraphics[width=0.85\textwidth]{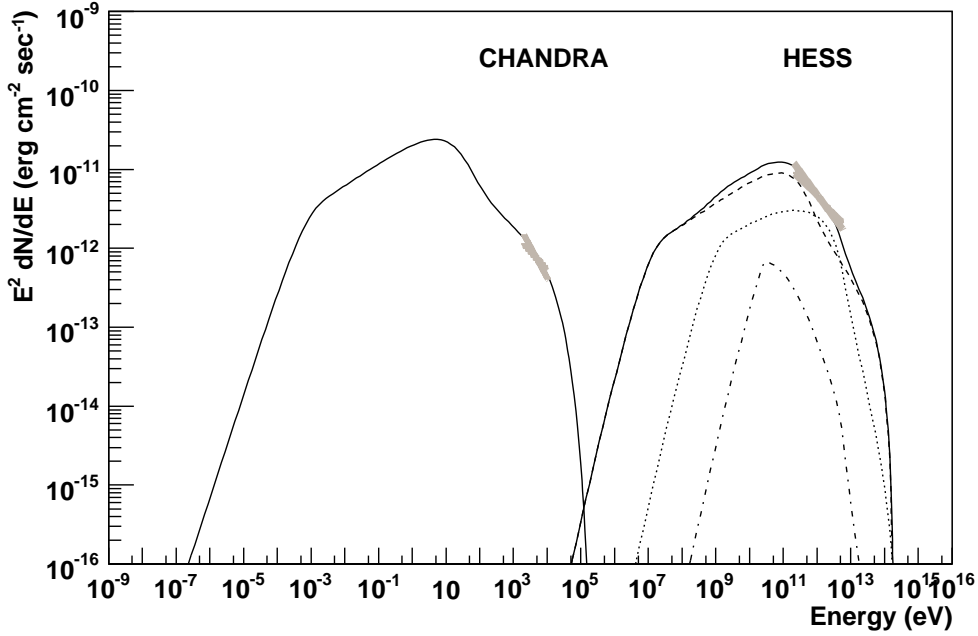}
\caption{\textit{Spectral energy distribution (SED) of G338.3-0.0
PWN emission, fitted with the one-zone time dependent model described
in the text. The VHE emission of HESS$~$J1640-465 together with the
total PWN X-ray spectrum derived with Chandra data are
shown. Parameters of this fit are summarized in Table 3. 
The IC emission (black line) is computed taking into account the CMB(dashed line), 
Star light (dotted line) and Dust (dotted and dashed line) targets, derived 
from Moskalenko et al.(2006).}}
\end{center}
\end{figure}


\end{document}